\documentclass[12pt,preprint]{aastex}

\shorttitle{GRB~030328 and GRB~041006 Chandra LETGS Spectra}
\shortauthors{Butler et al.}

\def\gtrsim{\mathrel{\hbox{\rlap{\hbox{\lower4pt\hbox{$\sim$}}}\hbox{$>$}}}}
\def\lessim{\mathrel{\hbox{\rlap{\hbox{\lower4pt\hbox{$\sim$}}}\hbox{$<$}}}}
\def\chandra{{\it Chandra}~}
\def\swift{{\it Swift}~}
\def\xmm{{\it XMM}~}
\def\hete{{\it HETE-2}~}

\begin{document}

\title{High Resolution Gratings Spectroscopy of GRBs 030382 and 041006 with
Chandra LETGS}
\author{
N.~R. Butler\altaffilmark{1}, 
G. R. Ricker\altaffilmark{1}, 
P. G. Ford\altaffilmark{1}, 
R. K. Vanderspek\altaffilmark{1},
H. L. Marshall\altaffilmark{1}, 
J. G. Jernigan\altaffilmark{2},
G. P. Garmire\altaffilmark{3},
and D. Q. Lamb\altaffilmark{4}.}
\altaffiltext{1}{Massachusetts Institute of Technology,
  Cambridge, MA 02139, USA}
\altaffiltext{2}{University of California at Berkeley, Space Sciences
  Laboratory, Berkeley, CA, 94720-7450}
\altaffiltext{3}{Department of Astronomy, Pennsylvania State University,
University Park, PA, 16802}
\altaffiltext{4}{Department of Astronomy and
  Astrophysics, University of Chicago, IL, 60637}

\begin{abstract}
We present high resolution X-ray spectroscopy of two recent 
GRB afterglows observed with the Low Energy Transmission Gratings on 
\chandra.   The afterglows to GRBs 030328 and 041006 are detected beginning
15.33 and 16.8 hours after each burst, respectively, and are
observed to fade in time during each $\sim 90$ ksec observation.
We fit for the continuum emission 
in each full data set and for the data sliced into half and quarter time 
sections.  For both afterglows, the continuum emission is well described 
by an absorbed power-law model, and the model parameters describing the 
absorption and spectral slope do not appear to evolve in time.  We perform 
a careful search for deviations from the model continua for the full and 
time-sliced data and find no evidence for significant ($\gtrsim 3\sigma$) 
narrow emission/absorption lines or edges.  The lack of detections
implies that line emission--if it is a general feature in GRB X-ray
afterglows--occurs early ($t\sim 0.3$ days in the source frame) and/or 
is short-lived ($dt \lessim 10$ ksec).  We also comment on synchrotron
fireball models for the X-ray and optical data.
\end{abstract}


\section{Introduction}
\label{sec:intro}

One of the great puzzles in the study of $\gamma$-ray bursts (GRBs) is that
of the X-ray afterglow lines.  Claims of low to moderate 
significance emission lines have been made based on data from several
missions: Fe lines have been detected in afterglow data from {\it ASCA}
\citep{yoshida99}, {\it Beppo-SAX} \citep{piro99,antonelli00}, and
{\it Chandra} \citep{piro00}; lines from highly ionized light, 
multiple-$\alpha$ elements like
Mg, Si, S, Ar, and Ca have been detected in afterglow data from {\it XMM}
\citep{reeves02,watson03} and {\it Chandra} \citep{butler03}.  The
detections are challenging to theorists because they typically imply large, 
concentrated masses of metals in the circumburst material 
\citep[see, e.g.,][]{lcg99} and a very efficient reprocessing of the 
non-thermal afterglow continuum into line radiation 
\citep[see,][]{bnr01,lrr02}.  For observers, the challenge is obtain
significant and unambiguous detections or firm upper limits.  
\citet{skr04} \citep[see also,][]{rutledge03} argue that the claims made
to date lack the necessary significance needed to prove that the X-ray 
lines are real.

For a burst at $z\sim 1$, the Low Energy Transmission Gratings
Spectrometer (LETGS) on \chandra \citep{weisskopf02} provides peak
sensitivity to H-like and He-like lines from Mg, Si, S, Ar, Ca, as
well as sensitivity to common transitions in Ni, Co, Fe, C, N, O, and 
Ne.  The spectral resolution ($E/\Delta E =10^2-10^3$ for $E=0.2-5$ keV) 
makes possible the fine centroiding
of lines and the determination of their shape.  This in turn, potentially
allows us to decide between different broadening mechanisms and to
distinguish line emission from narrow RRCs (radiative recombination
continua).  Ideally, we could then infer the chemical composition, 
kinematics, reprocessing mechanism (e.g. recombination or fluorescence), and 
even the geometry of the circumburst emitting medium.  Such were
the motivations that led us to
observe the afterglows to two recent GRBs detected by the {\it High Energy
Transient Explorer Satellite (HETE-2)}.  Both were bright GRBs,
in regions of the sky with little Galactic extinction.

\section{Observations}

The bright, long-duration GRB~030328 was detected by \hete at 11:20:58.34 UT,
with a $\gamma$-ray fluence of approximately 
$2.7 \times 10^{-5}$ erg cm$^{-2}$ \citep{joel03,atteia03}.   
\citet{pandp03} detected an uncatalogued optical source
within the \hete error region at 
R.A. $=12^h 10^m 48.4^s$, decl. $= -09^{\circ} 20' 51.3''$ (J2000.0).
\chandra acquired this target 15.33 hours after the burst and
observed with the LETGS/ACIS-S for 94 ksec (livetime) until 43.32 hours 
after the burst.  The mean counting rate for the X-ray afterglow
is 0.012 counts/s (summed over the dispersed signal from
the LETGS, and including the 0th order flux).  The
count rate over the full observation decays with a slope $\alpha=-1.5\pm 0.1$
($\chi^2/\nu = 34.3/30$, Figure \ref{fig:lightcurves}).

The bright, long-duration GRB~041006 was detected by \hete at 12:18:08 UT,
with a $\gamma$-ray fluence of approximately 
$7 \times 10^{-6}$ erg cm$^{-2}$ \citep{mark04}.  Early optical observations
by \citet{dacosta04} revealed an optical afterglow at the
coordinates:
 R.A. $=00^h 54^m 50.17^s$, decl. $= +01^{\circ} 14' 07.0''$ (J2000.0).
This source was observed with the LETGS/ACIS-S on \chandra for
86.3 ksec (livetime),
starting 16.8 hours after the burst and lasting until 42.57 hours after
the burst.  The coincident X-ray afterglow was detected with a mean 
counting rate of 0.017 counts/s.  The source faded in brightness
according to a power law, with a decay time slope of $\alpha=-1.0\pm 0.1$
($\chi^2/\nu = 34.5/28$).  The lightcurves for GRB~030328 and GRB~041006
are plotted with the concurrent observations in the optical in 
Figure \ref{fig:lightcurves}.

\section{Data Reduction and Continuum Fits}
\label{sec:continuum}

We reduce the LETGS spectral data from the L1 event lists using the
CIAO 3.2\footnote{http://cxc.harvard.edu/ciao/} processing tools.
We use version 2.28 of the calibration database (CALDB), which 
includes corrections for the low energy quantum efficiency
degradation.  We extract the $\pm1$ and 0th order LETGS spectra using
the standard scripts.  The 0th order data are extracted in circular
regions, and backgrounds are extracted outside of these regions with
care not to include counts from the dispersed spectra.
Spectral fitting and analysis is performed with 
ISIS\footnote{http://space.mit.edu/CXC/ISIS/}.  We fit the 0th order
and the combined $\pm1$ order data for each afterglow jointly.
The data are binned to a S/N $\ge 3.5$ per bin, and this restricts
the energy coverage to the 0.5-5.0 keV band.
We define S/N as the background-subtracted number of
counts divided by the square root of the sum of the signal counts and the
 variance in the background.  We fit each
 model by minimizing $\chi^2$.  All quoted errors are 90\% confidence.  

The GRB~030328 continuum is well fit ($\chi^2 = 61.34/63$) by a
an absorbed power-law with photon index $\Gamma = 2.0\pm 0.2$.  The
absorption column ($N_H=0.6^{+0.3}_{-0.2} \times 10^{21}$ cm$^{-2}$) 
is consistent with
the anticipated Galactic value in the source direction 
\citep[$N_H=10^{21}$ cm$^{-2}$;][]{dickey1990}.
This model implies an average 0.5-8 keV unabsorbed flux of $2.9\pm 0.2 \times
10^{-13}$ erg cm$^{-2}$ s$^{-1}$.  The data are less well fit 
($\chi^2 = 71.23/64$) by a thermal bremsstrahlung model with 
$kT=2.9^{+0.6}_{-0.5}$ keV and Galactic absorption.

The GRB~041006 continuum is also well fit ($\chi^2=80.50/83$) by an
absorbed power-law.  The absorbing column ($N_H=1.2^{+0.6}_{-0.5}
\times 10^{21}$ cm$^{-2}$) is larger than the Galactic value
\citep[$N_H=2.9 \times 10^{20}$ cm$^{-2}$;][]{dickey1990} 
at 99.9\% confidence
($\Delta \chi^2 = 10.3$, for 1 additional degree of freedom).
The best-fit photon number index is $\Gamma = 1.9\pm 0.2$.
The model flux (unabsorbed, 0.5-8 keV) is 
$4.8^{+0.4}_{-0.3} \times 10^{-13}$ erg cm$^{-2}$ s$^{-1}$.
The data are less well fit ($\chi^2=82.66/83$) by a thermal 
bremsstrahlung model with $kT = 4.8^{+1.7}_{-1.1}$ keV and
$N_H = 5\pm4 \times 10^{20}$ cm$^{-2}$.  

In order to test for spectral variability during each observation,
we divide the spectra into half and quarter time regions 
containing approximately equal numbers of counts.  As the number 
of counts per bin can be quite low--in violation of the assumption
which allowed us to employ $\chi^2$ fitting above--we minimize the
log of the Poisson probability $\exp{[-{\cal L}_{p0}/2]} \propto 
\prod_i (m_i+b_i)^{N_i }
\exp{[-(m_i+b_i)]} b_i^{B_i} \exp{[-r_i b_i]}$, where $N_i$ is the number
of source plus background counts in bin $i$, $m_i$ is the source
model evaluated for bin $i$, $b_i$ is the background model, 
$B_i$ is the off-source background, which is scaled by an area
factor $r_i$.  Rather than model the off-chip background,
we average over the background parameters $b_i$ above
for each spectral bin.   This averaging results in the fit statistic:
$${\cal L}_p(m|N,B) = 2 \sum_i \left \{ m_i - \log{ \sum_{n=0}^{n=N_i} 
{(n_i+B_i)! \over n!(N_i-n)!}
[m_i (1+r_i)]^{(N_i-n)} } \right \}. $$
We find that varying the model parameters which define the $m_i$ by minimizing
${\cal L}_p(m|N,B)$ yields best-fit values
and error regions for the full data sets which are consistent with those
found from $\chi^2$ fitting with background subtraction.  
Table 1 shows the results of our fits of the 
absorbed power-law models to the time-sliced data.  For each GRB afterglow,
the data are consistent with no spectral evolution.  

\section{Line Emission Upper Limits}
\label{sec:upperlims}

In Table 2, we report upper limits on the source
frame emission line equivalent widths and line fluxes for several common
ionic species.  The hosts of GRB~030328 and GRB~041006 have 
redshifts $z\ge 1.52$ \citep{martini03} and $z=0.712$ 
\citep{fugazza04}, respectively.  
We  consider Gaussian lines of width $\sigma \le 100$ eV in source frame.
We allow the line centroids to vary by 15\% from the
source frame value during the
fits, in order to allow for a possible blue- or redshift of the emitting
material with respect to
the burst source frame.  
The line associations are not intended to be unique, but only to span
the detector energy range with a minimum of redundancy.

We note that there is a marginal (3.1 $\sigma$, Table 2) 
detection of an O\_VIII line for GRB~030328.  We do not regard this line as
a serious candidate.  At a detected energy of 0.27 keV, the line is
present in only the $-1$ order of the LETGS.  We cannot confirm its
presence in the $+1$ order or the 0th order data.  (In the line
search below, we will not search out to such low energies.)  The line
is located on a region of the detector where the background dominates, 
and we cannot exclude the
possibility that it is due to a background fluctuation.  The implied 
luminosity and equivalent width are also quite large, and this would 
make it difficult to explain not detecting other emission lines.

\section{Line Search}
\label{sec:linesearch}

We search for emission and absorption lines in both
the full and time-sliced data over the 0.5-8 keV band.  
Assuming the model continua from 
Table 1, we examine the $\pm1$ order data 
at binnings $\Delta \lambda = $0.05,0.1,0.2,0.4,0.8 for deviations.
The finest binning here is approximately equal to the detector
resolution (FWHM).  We set a $4\sigma$ threshold for positive or
negative fluctuations.  A single bin deviation of $4\sigma$ 
correspond roughly to a $2\sigma$ detection in 500 trials.  Table 3
shows the 6 (5) candidates we find for
GRB~030328 (GRB~041006) as well as refined significance estimates
determined from fits of Gaussian lines to the combined 0th order
and $\pm1$ order data.  We find no highly significant features.

\section{Discussion}

\subsection{Lightcurves, Afterglow Synchrotron Modeling}

\begin{figure}[ht]
\centering
\includegraphics[width=4.5in]{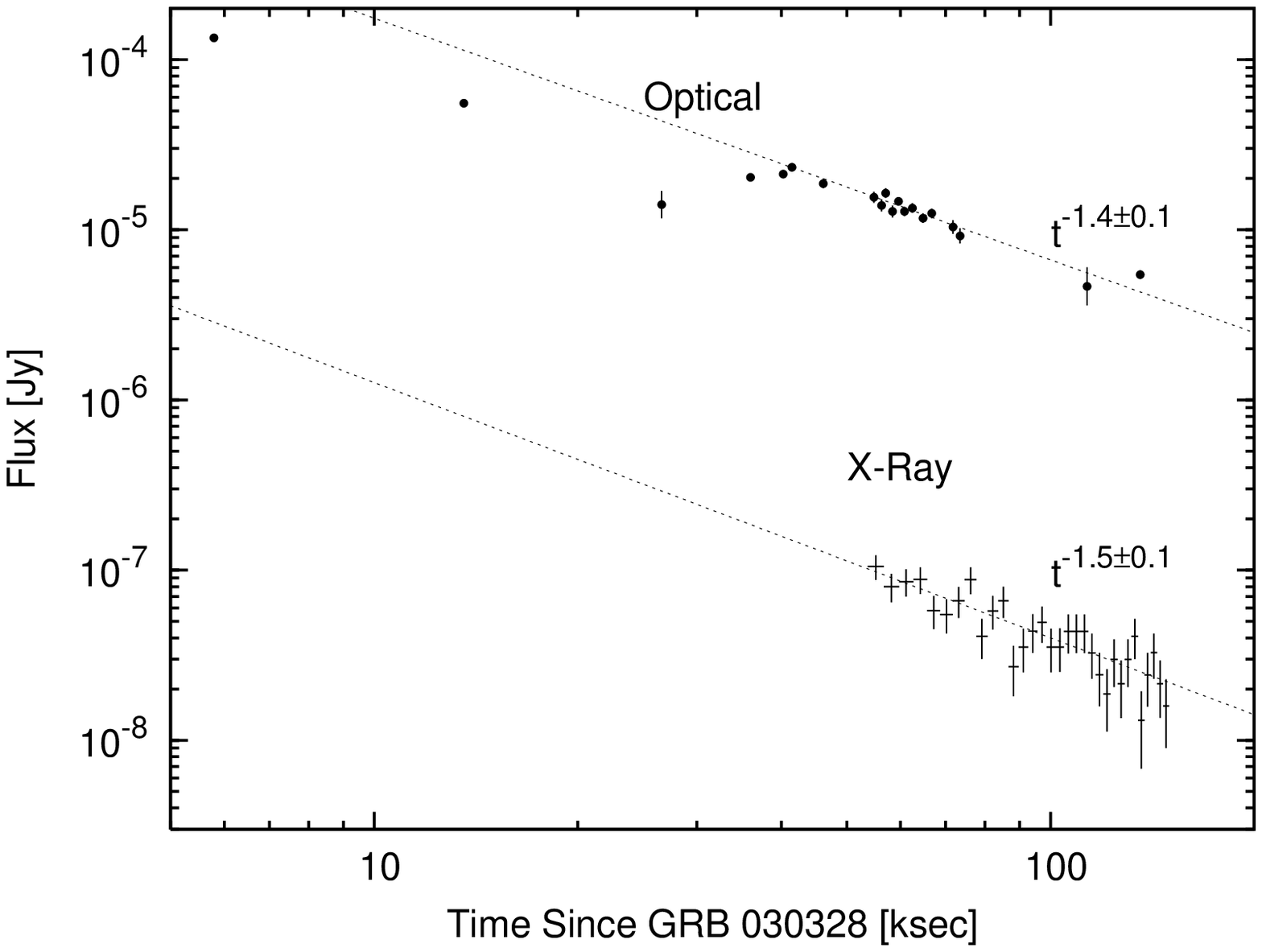}
\includegraphics[width=4.5in]{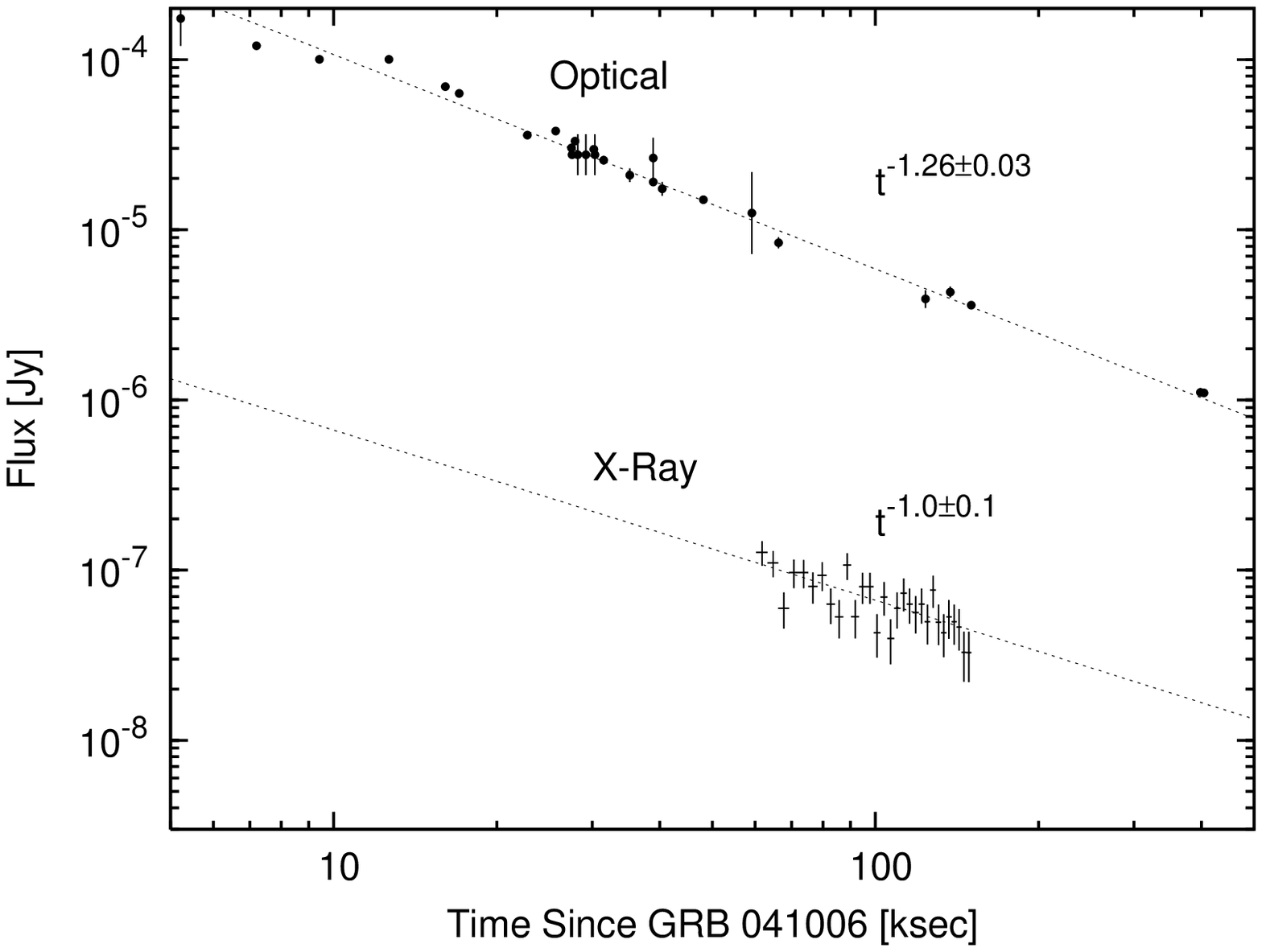}
\caption{
\footnotesize
The light curves for the afterglows to GRB~030328 (left) and GRB~041006,
in comparison with the optical.  (a) Optical data for GRB~030328 are from 
\citet{pandp03,fugazza03,galyam03,burenin03,rbp03,amj03,bartolini03,
gms03,ibrahimov03}.
(b) Optical data for GRB~041006 are from \citet{pdcn04,kahharov04,yost04,
ferrero04,ayaniyamaoka04,fugazza04,kinoshita04,
davanzo04,monfardini04,gzp04,mizrapandey04,rbp04,fynbo04,greco04,
kinugasatorii04,williams04}.  Corrections for Galactic extinction are
applied to the optical data from \citet{sfd98}.
}
\label{fig:lightcurves}
\end{figure}

The optical and X-ray lightcurves for GRB~030328 appear to fall
off at a consistent rate beginning $\sim t_{\rm GRB} + 40$ ksec 
(Figure \ref{fig:lightcurves}a).  
The ratio of fluxes implies a
broadband spectral slope $\beta_{\rm OX}=-0.8$, which is consistent
with the slope measured in the X-ray band, $\beta_{\rm X} = 1-\Gamma
 = -1.0\pm 0.2$ (Section \ref{sec:continuum}).  
If we associate the break at this
time as due to a collimated jet \citep[see, e.g.,][]{frail01}, we 
can derive the true $\gamma$-ray energy release from the GRB.
The relatively slow fade during the observation
is explained as due to a hard distribution of synchrotron
emitting electrons, with number index $p=1.6$.  
For $t_{\rm jet}=40\pm 5$ ksec, we find a jet opening angle of 
$\theta_{\rm jet}=2.^{\circ}8\pm 0.^{\circ}2$,
and and beaming-corrected GRB fluence of 
$E_{\gamma}=5.5^{+1.5}_{-1.2} \times 10^{50}$ erg.  
Here we use the formalism developed in \citet{sph99}, assuming a 20\% 
efficiency
for the conversion of kinetic energy into $\gamma$-rays and assuming
a uniform circumburst density of 3 cm$^{-3}$.  Here and throughout
we consider a flat cosmology with $H_{\circ}=70$ km s$^{-1}$ Mpc$^{-1}$
and $\Omega_{\Lambda}=0.7$.  

The value of $E_{\gamma}$ we derive is consistent with the value
derived in \citet{ghirl04}, and it is in agreement with the ``standard
energy'' of \citet{bfk03}.   However, if we calculate the isotropic X-ray 
luminosity at $t=10$ hours, $L_{\rm X,10}=1.6\pm0.4 \times 10^{43}$ erg 
s$^{-1}$, we find a number twenty times smaller than the X-ray standard
energy of \citet{bkf03}.  The X-ray standard energy would require 
$\theta_{\rm jet}\approx 12.^{\circ}5$ and $t_{\rm jet}\approx 24$ days
for this event.  If the X-ray afterglow luminosities do cluster in general
about a standard energy, the low flux may be telling us that 
the rapid light curve fade is due to a wind density profile and not a jet 
break.  Spherical expansion into a wind medium would yield temporal and 
spectral indices consistent with those measured for $p=2.4$ 
\citep{chevNli00}.
Although the $p=1.6$ found above is not uncommon in GRB afterglows 
\citep[see, e.g.,][]{pnk02},
a value of $p=2.4$ is favored for shock acceleration.

The lightcurve for GRB~041006 (Figure \ref{fig:lightcurves}b) also
shows a break at early times ($t=13\pm2$ ksec).  This
may not be a jet break, because the fades after the 
break are gradual ($t^{-1}$ rather than t$^{-2}$) and the energies
we would infer, $E_{\gamma}=2.9\pm0.7 \times 10^{49}$ erg
and $L_{\rm X,10}=2.2\pm0.6 \times 10^{42}$ erg s$^{-1}$ 
(for $\theta_{\rm jet}=3.^{\circ}2\pm0.^{\circ}2$), are orders of magnitudes
below the standard energies.
The spectral slope in the X-rays ($\beta_{\rm X} = -0.9\pm 0.2$;
Section \ref{sec:continuum}) is consistent with the slope measured in the
optical, $\beta_{\rm O} = -1.0 \pm 0.1$ \citep{gzp04,williams04}.
The broadband slope ($\beta_{\rm OX} = -0.7$) is consistent with the
X-ray slope, but it is more shallow than the optical slope.
The X-ray flux is six times higher than would be expected from an
extrapolation of the optical flux.  This suggests that the
X-ray spectrum is dominated by Inverse-Compton emission 
\citep{sariesin01}.
That mechanism--or possibly continued energy injection from the GRB 
source--could also explain the apparently slower X-ray than optical 
fade (Figure \ref{fig:lightcurves}b).

Host frame absorption may also be important, if
we have under-estimated the flux in the optical.  Because the
synchrotron cooling break may lie between the optical and X-ray bands, 
the unabsorbed optical flux may have as shallow a slope as
$\beta_{\rm O}\sim -0.4$.  Following \citet{galama01}, we determine that 
this change in slope can be accomplished
with a host-frame $A_V\approx 0.4$.  The X-ray data require a column density
in excess of the Galactic value (Section \ref{sec:continuum}).  
If we place the excess column
at the host, it is $N_H = 3.2\pm 0.16 \times 10^{21}$ cm$^{-2}$.  For
the Galactic $N_H-A_V$ relation \citep{pns95}, we would expect a corresponding
$A_V\approx 1.5$.  \citet{galama01} found that such a discrepancy is
common in GRB afterglows and may provide evidence for dust destruction.
Finally, we note that the late-time optical afterglow (at $t\gtrsim 5$ days)
is apparently brighter than the extrapolated flux shown in Figure
\ref{fig:lightcurves}b.  \citet{stanek05} present the late time data and
argue that the flux is dominated by an emerging supernova component.

\subsection{Sensitivity of the Line Search}

\begin{figure}[ht]
\centering
\resizebox{40pc}{!}{\includegraphics{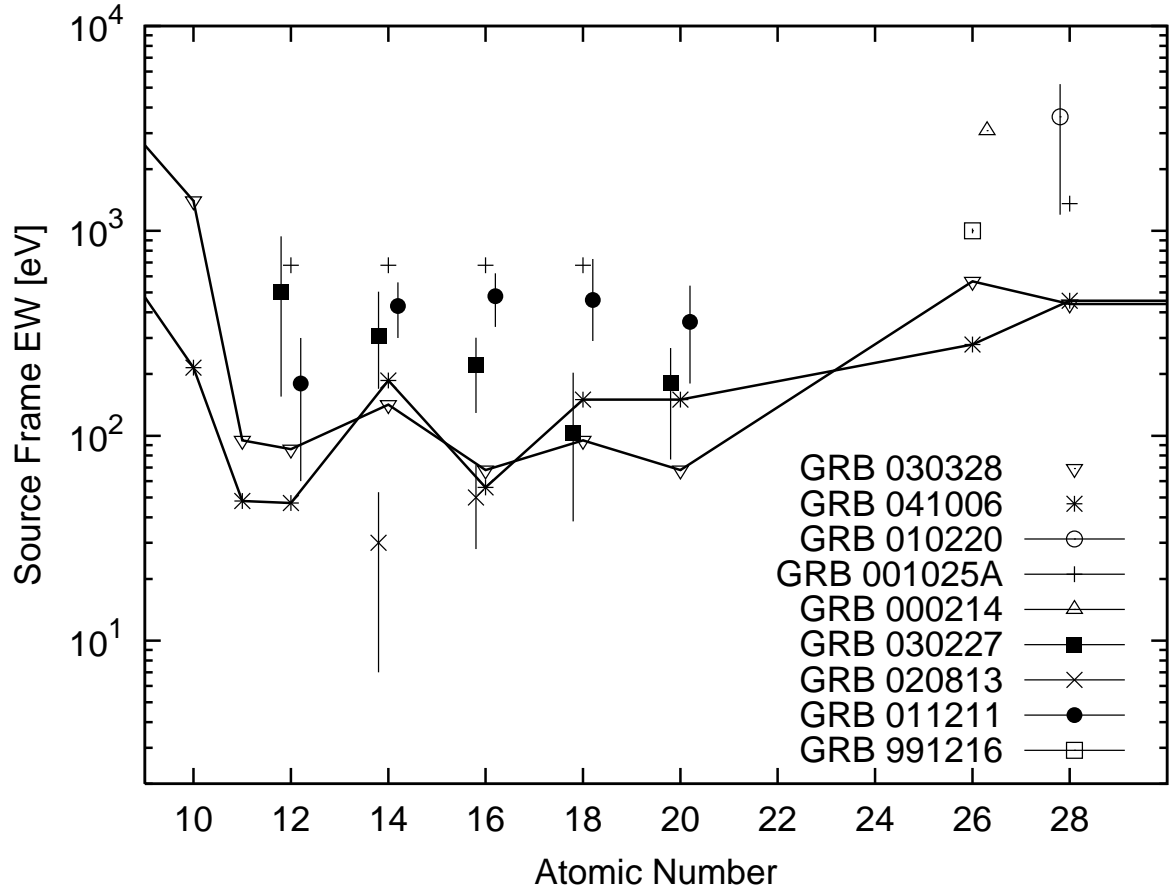}}
\caption{
\footnotesize
Comparison of equivalent width upper 90\% confidence upper limits for 
GRB~030328 (upward triangles) and GRB~041006 (asterisks) to the equivalent 
widths of emission lines claimed in the
literature \citep{butler03,reeves02,watson02,antonelli00,piro00,watson03}.
Error bars have been plotted where available.
}
\label{fig:eqwidth}
\end{figure}

Figure \ref{fig:eqwidth} shows how the 90\% confidence limits we derive for the
line equivalent widths compare with those derived for emission lines
claimed in the literature.  If the claimed lines from the
other observations had been present in our data for the full observations,
we estimate that we would have detected all except for the faint
GRB~020813 lines \citep{butler03}.  Because the lines claimed in \xmm data
for several events (GRB~011211, \citet{reeves02}; GRB~001025A,
\citet{watson02}; GRB~030227, \citet{watson03}) persisted
for extremely short periods ($10-30$ times
shorter than the GRB~030328 or GRB~041006 observations, viewed in 
the host frame), our limits
are not tight enough to rule them out.  The Fe line observed in only
$9.7$ ksec of \chandra HETGS data for a very bright GRB~991216 afterglow,
would also likely not be detected in these LETGS observations.

\begin{figure}[ht]
\centering
\resizebox{40pc}{!}{\includegraphics{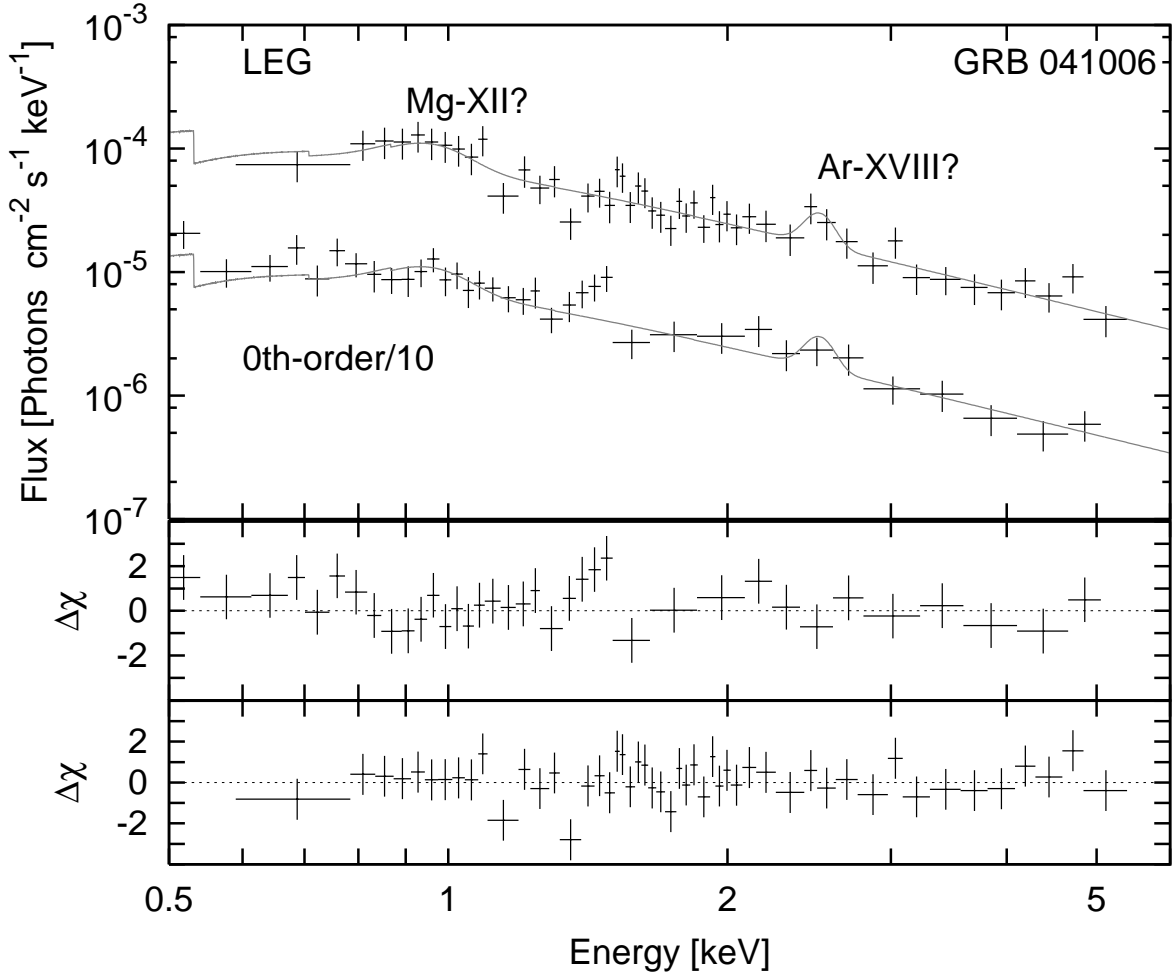}}
\caption{
\footnotesize
The most significant evidence found for emission lines in the GRB~030328 or
GRB~041006 spectra.
The addition of two broad ($\sigma_E=0.09$ keV) emission 
lines to the GRB~041006 spectrum for the full observation improves
the absorbed power-law fit at $2.9\sigma$ significance.
}
\label{fig:41006_flux2}
\end{figure}

In the case of the lines detected in the \xmm data for GRB~011211 and
GRB~030227, the individual lines are not significant ($\lessim 2\sigma$),
whereas the juxtaposition of multiple lines is moderately significant
($\gtrsim 3\sigma$).
We explore the possibility of significant sets of emission lines
by fitting for combinations of the candidates lines from Tables 2 and
3, for each time slice.  We only consider initial energies from ions
in Table 3 where a possible detection is better than 90\% confidence
(significance $>1.6\sigma$).  For the full data set, the addition of
two of the candidate lines
for GRB~041006 appears to improve the power-law fit at moderate
significance ($2.9\sigma$, $\Delta \chi^2 = 19.1$, for 6 additional
degrees of freedom).   The lines would be associated with H-like Mg and
Ar at $z=0.45\pm0.05$, requiring a blue-shift of $0.16\pm0.04$c from the host
at $z=0.712$.
Additional lines do not improve the fit
markedly.  We find no evidence for additional significant line sets in
the GRB~030328 or GRB~041006 full or time-sliced spectra.  Due to the
broadness of the lines in Figure \ref{fig:41006_flux2} ($\sigma_E=0.09$,
best fit), it is difficult to say
whether the lines are real or whether the continuum is not adequately
modelled.  Such a degeneracy may also exist for the
\xmm multiple-line claims (GRB~0001025A, \citet{watson02}, GRB~011211,
\citet{reeves02}; GRB~030227, \citet{watson03}), although the
significance appears to be greater in those cases.  In \citet{butler05}, 
we show
that the significance of the GRB~011211 lines depends strongly on 
whether or not the column density can be fixed at the Galactic value.
For the \xmm lines, we do not have the spectral resolution 
necessary to measure the breadth of the claimed lines in order to unambiguously
separate the line emission from the continuum emission.

\subsection{Comparison with Theoretical Predictions}
\label{sec:modpred}

\begin{figure}[ht]
\centering
\resizebox{40pc}{!}{\includegraphics{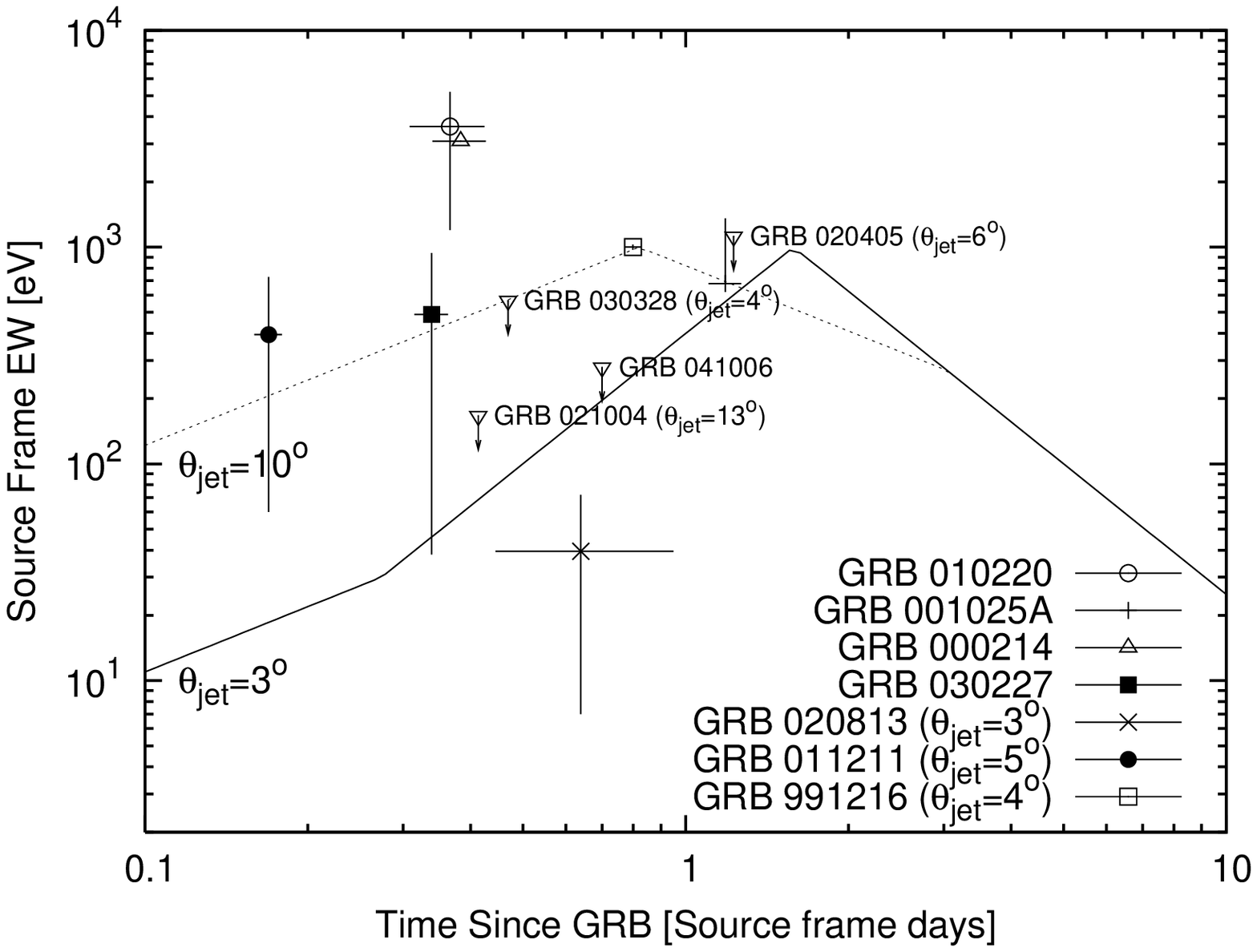}}
\caption{
\footnotesize
Comparison of measured $EW$ values with model predictions (Section \ref{sec:modpred}).  
We plot the equivalent width upper limits (downward arrows) for the
bursts studied here and for GRB~021004 \citep[{\it Chandra/HETGS};][]{butler03}
and GRB~020405 \citep[{\it Chandra/LETGS};][]{mirabal03}.
Also plotted are the $EW$ values for the emission lines claimed in the
literature \citep{butler03,reeves02,watson02,antonelli00,piro00,watson03}. 
Jet opening angles $\theta_{\rm jet}$ are taken from \citet{bfk03,ghirl04}.
}
\label{fig:eqwidthVtime}
\end{figure}

The afterglow line emission (if it is real) is thought to be due to 
photoionization on the surface of
optically thick slabs of metal-rich material surrounding the progenitor.
For the ``distant reprocessor'' models 
\citep[e.g., a ``supranova;''][]{vietri99}, material from a precursor
supernova
(at $R\sim 10^{16}$ cm) is excited by reflection of the afterglow 
continuum radiation.
In the ``nearby reprocessor'' scenario, a long lived
central engine \citep{rnm00} or a plasma bubble 
GRB \citep{mnr01} ionizes material on the sides of a funnel
carved out of the progenitor star by the GRB (at $R\sim 10^{13}$ cm).
\citet{bnr01} study the equivalent widths of Fe lines, for each
scenario, as a function of the continuum luminosity.  Roughly,
they find that the the equivalent widths peak near 1 keV in each scenario for 
solar abundances and incident
X-ray continuum luminosities of order $10^{46}$ erg s$^{-1}$.  
Because the reflected continuum plus line emission must compete
with the afterglow continuum emission at the observation epoch,
this sets an upper limit on the observed equivalent widths.
\citet{gmk04} employ a toy model based on this behavior to 
estimate the possible significance of future detections at various 
observation times for \swift, \chandra, and \xmm.  Here, we perform
a similar exercise in order to better understand the GRB~030328 and 
GRB~041006 emission line upper-limits in the context of the previously
claimed detections (Figure \ref{fig:eqwidthVtime}).

We assume that the equivalent widths increase
to 1 keV with decreasing luminosity as $EW\propto 1/L_{\rm X}$, then
decrease below $10^{46}$ erg s$^{-1}$ as $EW\propto L_{\rm X}$.  For
$L_{\rm X}$ we assume the standard energy of \citet{bkf03} at $t=10$ hours
in the source frame.  This decreases as $t^{-1}$ until the jet break
at $t_{\rm jet}$, whence the flux starts decreasing more rapidly as
$t^{-2}$.  We account for the diversity in observed
continuum fluxes and break times $t_{\rm jet}$ in terms of a range of jet opening 
angles $\theta_{\rm jet}$, where we set 
$t_{\rm jet}=(\theta_{\rm jet}/0.1)^2$.  Two of the resulting $EW(t)$
models are shown in Figure \ref{fig:eqwidthVtime}.  At a given redshift,
the narrowly beamed event would be observed to have a higher continuum flux 
due to the increased beaming.  In Figure \ref{fig:eqwidthVtime}, we plot
the Fe line equivalent widths derived here and from other \chandra observations.
We also plot the equivalent widths of detected light metal and Fe lines quoted in the 
literature.  The fluxes from the light metal
lines can equal the Fe line flux when the reflecting material is less ionized
\citep{lrr02}.  A similar $EW(L)$ relation arises if the
light metal line emission systematically arises at smaller radii than does
the Fe line
emission, as suggested by \citet{lrr02}.

Aside from the $EW$ values for GRB~020813 and GRB~030227, the detected line $EW$ 
values lie well above the model predictions.  The events with measured $\theta_{\rm jet}$
values are narrowly jetted events, for which the $EW$ at early times is expected to
be low.  The $EW$ values can be larger by a factor of
ten or so at early times if the photoionization in the nearby reprocessor
scenario occurs for very shallow incidence angles \citep{kmr03} or if
the reflecting material has ten times solar abundances
\citep[as for GRB~011211;][]{reeves02}.  Such mechanisms must be invoked to 
justify the line claims.  This is not necessarily the case, however, for
the Fe line upper limits derived here and for GRB~020205 
\citep[{\it Chandra/LETGS};][]{mirabal03} and  GRB~021004
\citep[{\it Chandra/HETGS};][]{butler03}.  These events apparently span a
broad range of jet opening angles from $\theta_{\rm jet}=4^{\circ}$ to
$\theta_{\rm jet}=13^{\circ}$ (Figure \ref{fig:eqwidthVtime}).  We can conclude
from this that potentially unusual conditions in the line-emitting material
(e.g. supersolar abundances) are likely as important or more important for line
production than is the strength of the continuum at the observation epoch.
Finally, although selection
on bright and slowly fading bursts would tend to select possibly less interesting
low-redshift events, the potentially high $EW$ values make them appealing.
Early observations of many such events by \swift may constitute our best hope
for building up statistics and
for solving the mystery of X-ray afterglow line emission.

\section{Conclusions}

Beyond a power-law decay in flux versus time, the \chandra LETGS spectra
for the X-ray afterglows to GRBs 030328 and 041006 do not appear to evolve 
in time.  We find little evidence for discrete emission features--emission or
absorption lines, narrow recombination edges, etc.--in these two high 
resolutions gratings observations.

The number of sensitive non-detections of line emission in gratings 
observations with \chandra is
growing.  If line emission is a general feature in GRB X-ray afterglows,
then the LETGS observations of GRB~030328 and GRB~041006 discussed 
here--alongside the LETGS observation of GRB~020405 
\citep{mirabal03} and the HETGS observation of GRB~021004 
\citep{butler03}--imply that the emission must occur early (prior
to $t\sim 0.3$ days in the source frame) and/or be short-lived ($dt \lessim
10$ ksec).
Early emission would favor the nearby reprocessor scenario and 
one-step explosions \citep[e.g. a hypernova;][]{woosley93}.
Sporadic or short-lived emission would imply a persistent and erratic
central engine or a clumpy circumburst medium.

\acknowledgments
We thank Harvey Tananbaum for his generous allocation of Director's 
Discretion Time for the GRB~030328 observation.
This research was supported in part by NASA contract NASW-4690.

\clearpage
\begin{table}
\begin{center}
\caption{Time-resolved Spectroscopy}
\vspace{5mm}
\begin{tabular}{cccc}\hline\hline
 GRB~030328 & & & \\\hline
 & $N_H$ ($10^{21}$ cm$^{-2}$) & $\Gamma$ & Time Coverage (ksec) \\\hline
half 1    & $0.5\pm 0.3$        & $2.0\pm 0.2$        & 00.0-29.8  \\
half 2    & $0.7^{+0.6}_{-0.4}$ & $2.0\pm 0.2$        & 29.8-92.7  \\
quarter 1 & $0.9^{+0.9}_{-0.6}$ & $2.1^{+0.4}_{-0.3}$ & 00.0-12.4  \\
quarter 2 & $0.3^{+0.4}_{-0.3}$ & $1.9^{+0.3}_{-0.2}$ & 12.4-29.8  \\
quarter 3 & $0.3^{+0.6}_{-0.3}$ & $1.9^{+0.3}_{-0.2}$ & 29.8-55.4  \\
quarter 4 & $1.2^{+1.0}_{-0.7}$ & $2.2^{+0.4}_{-0.3}$ & 55.4-92.7  \\\hline
 GRB~041006 & & & \\\hline
 & $N_H$ ($10^{21}$ cm$^{-2}$) & $\Gamma$ & Time Coverage (ksec) \\\hline
half 1    & $1.3^{+0.6}_{-0.5}$ & $1.9\pm 0.2$ & 00.0-33.8  \\
half 2    & $1.4\pm 0.6$        & $2.0\pm 0.2$ & 33.8-86.3  \\
quarter 1 & $1.7^{+0.8}_{-0.7}$ & $2.1\pm 0.3$ & 00.0-14.9  \\
quarter 2 & $1.2^{+1.0}_{-0.8}$ & $1.8\pm 0.3$ & 14.9-33.8 \\
quarter 3 & $0.9^{+0.9}_{-0.7}$ & $1.8\pm 0.3$        & 33.8-57.6 \\
quarter 4 & $2.1\pm 1.0       $ & $2.2^{+0.4}_{-0.3}$ & 57.6-86.3 \\\hline
\end{tabular}
\end{center}
\label{tab:time_resolve}
\end{table}

\begin{table}
\begin{center}
\caption{Emission Line Upper Limits}
\vspace{5mm}
\begin{tabular}{ccccc}\hline\hline
GRB~030328 & & & & \\\hline
  Ion  &     $E_{\rm line,rest}$  (keV) &    $L_{\rm line}$ ($10^{43}$ erg s$^{-1}$) &  EW (eV) &    Signif. \\\hline
Ni\_XXVIII & 8.073 & $\le 3.9 $ & $\le 440 $ & $\le 0.7\sigma $ \\
Fe\_XXVI & 6.952 & $\le 3.8 $ & $\le 367 $ & $\le 1.0\sigma $ \\
Fe\_XXIV & 6.400 & $\le 6.8 $ & $\le 567 $ & $\le 1.3\sigma $ \\
Ca\_XX & 4.100 & $\le 1.3 $ & $\le 68 $ & $\le 0.6\sigma $ \\
Ar\_XVIII & 3.318 & $\le 2.1 $ & $\le 95 $ & $\le 1.4\sigma $ \\
S\_XVI & 2.612 & $\le 1.9 $ & $\le 68 $ & $\le 0.9\sigma $ \\
Si\_XIV & 2.000 & $\le 5.2 $ & $\le 142 $ & $\le 0.9\sigma $ \\
Mg\_XII & 1.472 & $\le 4.3 $ & $\le 86 $ & $\le 1.9\sigma $ \\
Na\_XI & 1.236 & $\le 5.7 $ & $\le 95 $ & $\le 0.1\sigma $ \\
Ne\_X & 1.022 & $\le 1.8 \times 10^2 $ & $\le 1.4 \times 10^3 $ & $\le 1.5\sigma $ \\
O\_VIII & 0.653 & $\le 9.0 \times 10^2 $ & $\le 4.9 \times 10^3 $ & $\le 3.1\sigma $ \\\hline
GRB~041006 & & & & \\\hline 
  Ion  &     $E_{\rm line,rest}$  (keV) &    $L_{\rm line}$ ($10^{43}$ erg s$^{-1}$) &  EW (eV) &    Signif. \\\hline
Ni\_XXVIII & 8.073 & $\le 1.5 $ & $\le 456 $ & $\le 1.0\sigma $ \\
Fe\_XXVI & 6.952 & $\le 1.1 $ & $\le 279 $ & $\le 0.9\sigma $ \\
Fe\_XXIV & 6.400 & $\le 0.9 $ & $\le 196 $ & $\le 0.4\sigma $ \\
Ca\_XX & 4.100 & $\le 1.1 $ & $\le 150 $ & $\le 1.6\sigma $ \\
Ar\_XVIII & 3.318 & $\le 1.1 $ & $\le 150 $ & $\le 1.6\sigma $ \\
S\_XVI & 2.612 & $\le 0.6 $ & $\le 56 $ & $\le 2.0\sigma $ \\
Si\_XIV & 2.000 & $\le 2.6 $ & $\le 186 $ & $\le 2.0\sigma $ \\
Mg\_XII & 1.472 & $\le 0.9 $ & $\le 47 $ & $\le 1.6\sigma $ \\
Na\_XI & 1.236 & $\le 1.1 $ & $\le 48 $ & $\le 1.4\sigma $ \\
Ne\_X & 1.022 & $\le 7.7 $ & $\le 215 $ & $\le 0.6\sigma $ \\
O\_VIII & 0.653 & $\le 39.9 $ & $\le 1049 $ & $\le 0.4\sigma $ \\
N\_VII & 0.500 & $\le 7.5 \times 10^3 $ & $\le 1.1 \times 10^5 $ & $\le 2.4\sigma $ \\\hline
\end{tabular}
\end{center}
{\small
Note.---Rest-frame flux ($L_{\rm line}$) and source frame equivalent width (EW) upper limits are on $L_{\rm line}$ and EW are
90\% confidence.
}
\label{table:upperlim}
\end{table}

\begin{table}
\begin{center}
\caption{Emission/Absorption Line Candidates}
\vspace{5mm}
\begin{tabular}{ccccccc}\hline\hline
GRB~030328 & & & & & \\\hline
$E_{\rm line}$  & Assoc. & z &	Signif.	& EW  & 	$L_{\rm line}$  & Region \\
 (keV) &  & &	& (keV) &  ($10^{43}$ erg s$^{-1}$) &  \\\hline
2.345  & Fe\_XXIV  & 1.729 &  $2.0\sigma$  &   290  &   2.5   &  half2  \\
0.561  & Mg\_XII   & 1.623 &  $1.7\sigma$  &    90  &   3.3   &  half2  \\
0.727  & Si\_XIV   & 1.751 &  $2.6\sigma$  &   180  &   17.0  &  qtr1   \\
2.389  & Fe\_XXIV  & 1.679 &  $1.9\sigma$  &   430  &   4.5   &  qtr3   \\
0.721  & Si\_XIV   & 1.774 &  $2.2\sigma$  &   100  &   2.8   &  qtr3   \\
0.538  & Mg\_XII   & 1.736 &  $1.6\sigma$  &   240  &   10.0  &  qtr3   \\\hline
GRB~041006 &  & & & & \\\hline
$E_{\rm line}$  & Assoc. & z &	Signif.	& EW  & 	$L_{\rm line}$  & Region \\
 (keV) &  & & 	& (keV) &  ($10^{43}$ erg s$^{-1}$) &  \\\hline
0.993 &	Mg\_XII   & 0.482 & $2.0\sigma$ & 110 &	1.6 &	full \\
2.300 & Ca\_XX	  & 0.783 & $2.1\sigma$ & 170 &	0.8 &	half2 \\
2.146 &	Ar\_XVIII & 0.546 & $2.1\sigma$ & 240 &	3.0 &	qtr1 \\
0.558 &	Ne\_X	  & 0.832 & $1.9\sigma$ & 210 &	6.4 &	qtr2 \\
0.543 &	Ne\_X	  & 0.882 & $1.4\sigma$ & 170 &	3.3 &	qtr3 \\\hline
\end{tabular}
\end{center}
{\small
Note.---Significances (``Signif.'') estimated from $\Delta {\cal L}_p$, assuming a $\chi^2$ distribution with $\nu =3$.
}
\label{table:candidates}
\end{table}

\clearpage

\begin{figure}[ht]
\centering
\resizebox{40pc}{!}{\includegraphics{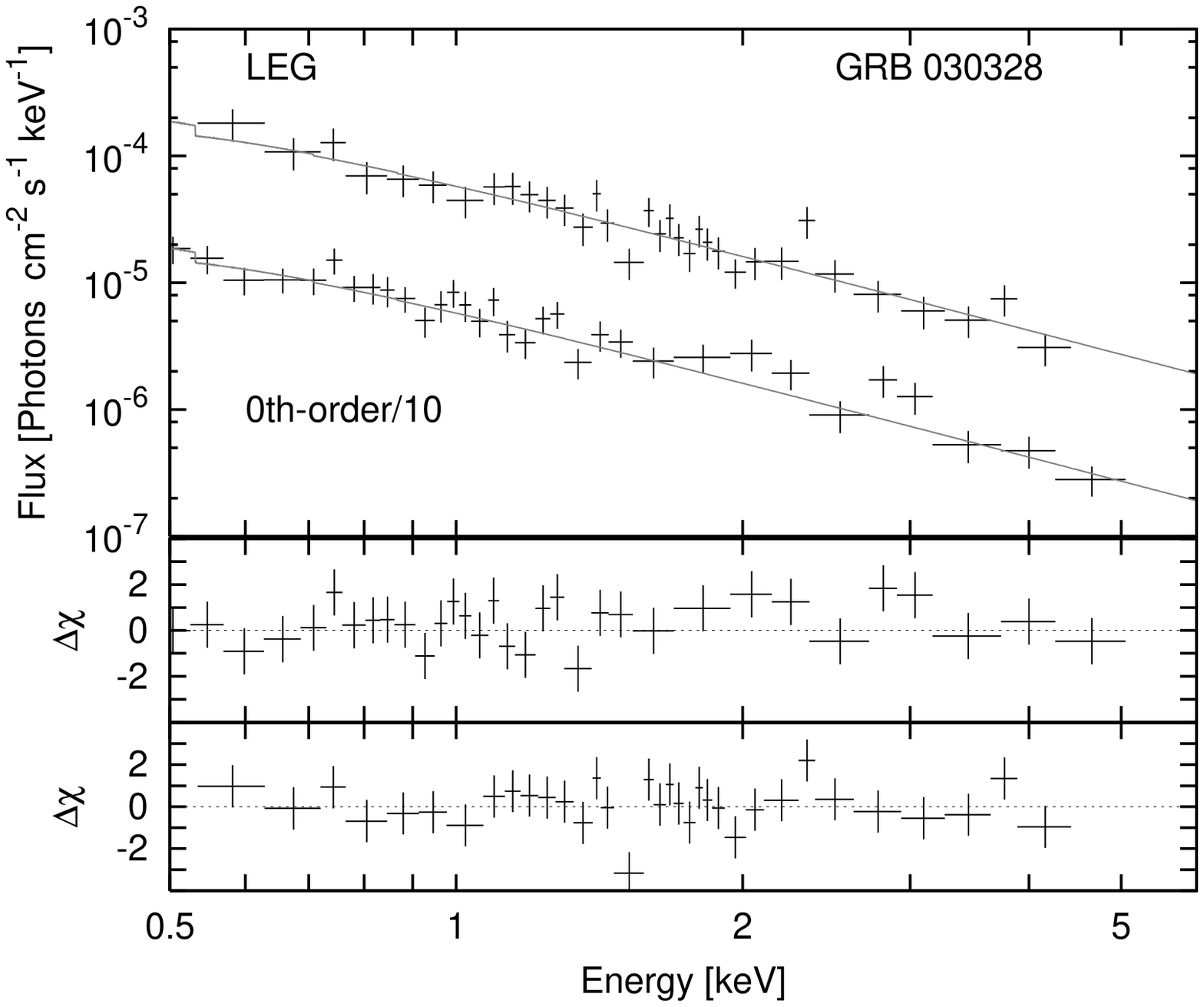}}
\caption{
\footnotesize
Combined-first and 0th-order spectra for GRB~030328, fit with an absorbed power-law.
}
\label{fig:30328_flux}
\end{figure}

\begin{figure}[ht]
\centering
\resizebox{40pc}{!}{\includegraphics{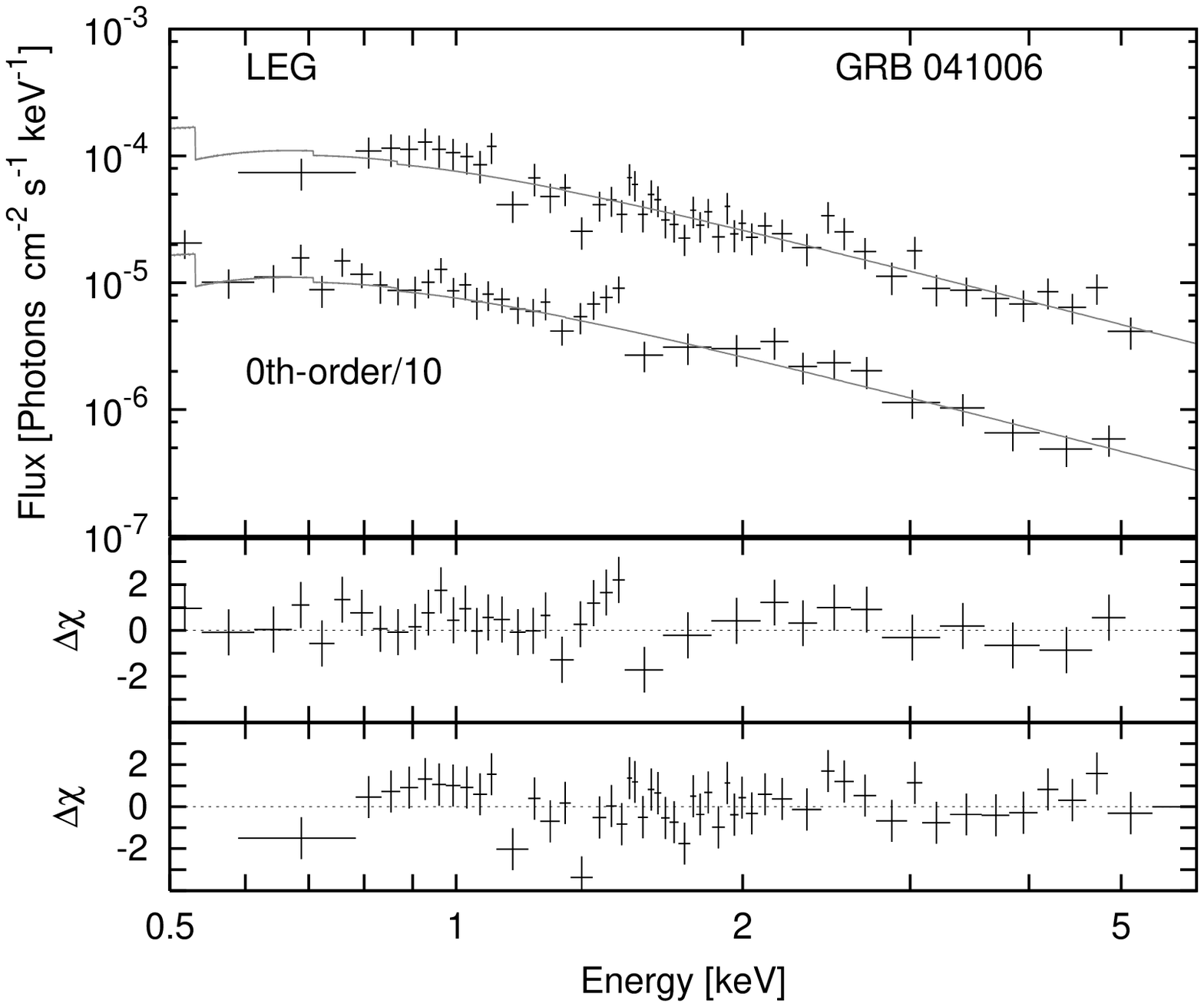}}
\caption{
\footnotesize
Combined-first and 0th-order spectra for GRB~041006, fit with an absorbed power-law.
}
\label{fig:41006_flux}
\end{figure}

\end{document}